\documentclass[12pt,amsfonts]{article}    
\usepackage{amsfonts}

\def\one{1\hskip-.37em 1}

\def\ir{{\rm I}\hskip-.2em{\rm R}}
\def\ts{\textstyle}
\def\vp{\varphi}
\def\bl{\bf{\l}}
\def\mfh{\mathfrak h}
\def\half{\textstyle{\frac{1}{2}}}
\def\threequarter{\textstyle{\frac{3}{4}}}
\def\halfnu{\textstyle{\frac{1}{2\nu}}}
\def\iN{{\rm I}\hskip-.2em{\rm N}}
\def\iH{{\rm I}\hskip-.2em{\rm H}}
\def\H{{\cal H}}
\def\ZZ{\cal Z}
\def\mfH{\mathfrak{H}}
\def\threebytwo{\textstyle{\frac{3}{2}}}
\def\eigth{\textstyle{\frac{1}{8}}}
\def\EE{{\cal E}}
\def\p{\phi}

\def\th{\theta}
\def\H{{\cal H}}
\def\B{\beta}
\def\v{\vskip.3cm}
\def\g{\gamma}
\def\kk{k^{\sf plus}}
\def\l{\lambda}
\def\D{{\cal D}}
\def\si{\sigma}
\def\S{\Sigma'}
\def\t{\textstyle}
\def\te{}  
\def\De{\Delta}
\def\F{{\cal F}}
\def\tr{\rm Tr}
\def\E{{\rm I}\hskip-.2em{\rm E}}
\def\ra{\rightarrow}
\def\tint{{\textstyle\int}}
\def\hg{{\hat g}}
\def\hp{{\hat\pi}}
\def\hph{{\hat\phi}}
\def\s{\hskip.08em}
\def\P{\Pi'}
\def\d{\partial}
\def\o{\overline}
\def\a{\alpha}
\def\b{\begin{eqnarray*}}  
\def\e{\end{eqnarray*}}    
\def\bn{\begin{eqnarray}}  
\def\en{\end{eqnarray}}   

\def\<{\langle}
\def\>{\rangle}
\def\g{\gamma}
\def\bk{\mathbf k}
\def\bm{\mathbf m}
\def\dn{d^n\!x}
\def\no{\nonumber}

\def\ds{d^s\!x}
\def\k{\kappa}
\def\bl{\bold l}
\def\quarter{\textstyle{\frac{1}{4}}}

\def\hk{\hat{\kappa}}
\def\{{\lbrace}
\def\}{\rbrace}
\begin{document}
\title{Building a Genuine Quantum Gravity}
\author{John R. Klauder\footnote{john.klauder@gmail.com} \\
Department of Physics and Department of Mathematics \\
University of Florida,   
Gainesville, FL 32611-8440}
\date{ }
\bibliographystyle{unsrt}
\maketitle
\begin{abstract}
An affine quantization approach leads to a genuine quantum theory of general relativity by extracting insights from a short list of increasingly more complex, soluble, perturbably nonrenormalizable models.
\end{abstract}

\section{Guidance to the Quantization of \\Nonrenormalizable Models}
The rules of canonical quantization for the all-important Hamiltonian operator allow for added $\hbar$-counterterms based on `experimental determination'.
Sometimes this means that different physical systems may need different counterterms, which is readily illustrated
 by the elementary example that the simple harmonic oscillator has a positive zero-point energy while as one of the modes of a free scalar field it has no zero-point energy. Normal ordering proves to be a satisfactory renormalization for a number of field theories, but for
 perturbably nonrenormalizable models it
fails significantly. Are there other $\hbar$-counterterms that lead to acceptable results for nonrenormalizable models?  We address this question and seek to find alternative $\hbar$-additions that lead to acceptable results.  In so doing, we study this question with the help of affine field quantization,
a procedure that is largely reviewed in the following section, and we learn that this program offers a suitable counterterm for
perturbably nonrenormalizable ultralocal models,  which may well be
suitably modified for more complicated models such as perturbably nonrenormalizable covariant scalar models in 5 and more spacetime dimensions, and also
  perturbably  nonrenormalizable general relativity in 4 spacetime dimensions.

As an initial comment, we recall that if a classical model is perturbably nonrenormalizable when conventionally quantized, then in both the
classical and the quantum realms, the interacting theories are
{\it not} continuously connected to their conventional free theories when the interaction is reduced to zero. As a simple
`perturbably nonrenormalizable toy model', consider the action functional {\it (and its associated domains!)} for $y\in\mathbb{R}$ and $g\ge0$, given by
   \bn  A_g=\tint_0^T \{\half[ \dot{y}(t)^2- \omega^2\s y(t)^2]-g\s y(t)^{-4}\s\}\,dt \;, \en
 which illustrates the point clearly when initially $g>0$ and then $g\ra0$. The result is {\it not} the free theory classically or quantum mechanically, but instead provides an example of what we call a `pseudofree model' (e.g., see \cite{jrk1}).

\section{Ultralocal Models}
\subsection{Canonical Quantization}
The classical Hamiltonian for this model is given by
  \bn H(\pi,\phi) =\tint\{\half[\s\pi(x)^2+m_0^2\s\phi(x)^2]+g_0\s \phi(x)^4\s\}\,d^s\!x\;, \en
 where the momentum field $\pi(x)$ and the scalar field $\phi(x)$ have a Poisson bracket $\{\phi(x),\s\pi(x')\}=\delta^s(x-x')$, and $x\in\mathbb{R}^s$ is a point in an $s$-dimensional configuration space, $s\ge1$,
 \cite{JrK}. 
 Conventional canonical quantization leads to a {\it free} field solution in which $m_0\ra m$ and $g_0\ra0$.
 Briefly reviewed, the conventional approach begins with (i) a regularization in which a finite subspace of $\mathbb{R}^s$ is replaced with a finite, discrete spatial lattice
 of points $\bk\s a$, where $\bk=\{k_1,k_2,\ldots,k_s\}$, $k_j\in\mathbb{Z}=\{0,\pm1,\pm2,\cdots\}$, $j=1,\cdots,s$, composed of $N'<\infty$ points, with $a>0$ as the lattice spacing, (ii) a quantization of the regularized system, followed by (iii) an elimination of the regularization as $a\ra0$. As dictated by the Central Limit Theorem, the result is  a Gaussian ground state, i.e., {\it a free
 quantum system}, in which $g_0=0$. The classical limit of the quantum theory is also free and thus contradicts the original, nonlinear
 classical theory.

 We aim to do better. The story of the first example, as presented below, is offered in more detail than the other examples because the
 other examples have rather similar stories.

\subsection{Affine Quantization}
 An affine quantization starts by first introducing the classical affine field $\k(x)\equiv\pi(x)\phi(x)$, $\phi(x)\ne0$,  and the affine field replaces the momentum field. The Poisson bracket is $\{\phi(x),\s\k(x')\}=\delta^s(x-x')\s\phi(x)$, $\phi(x)\ne0$, which can lead to a representation
  of the affine Lie algebra. Now the classical Hamiltonian is given by
     \bn  H'(\k,\phi)=\tint\{ \half[\k(x)\phi(x)^{-2}\k(x)+m_0^2\s\phi(x)^2]+g_0\s\phi(x)^4\}\, d^s\!x\;. \en
  The quantum commutator is given by $[\hph(x),\s\hk(x')]=i\hbar\s\delta^s(x-x')\s\hph(x)$, $\hph(x)\ne0$, and the affine operators have two irreducible representations: one where $\hph(x)>0$ and one where $\hph(x)<0$. Thus, for the affine field $\hph(x)$, we will use these two irreducible representations, a possibility offered by the rules of Enhanced Quantization pertaining to reducible operator representations
  \cite{jrkEQ}. 

   The commutator resembles a current commutation relation, and, as such, we find the field operators have a different kind of representation suitable for operator product expansions. Let $\l\in{\mathbb{R}}$, then $\hph(x)=\tint B(x,\l)^\dag\s\l\s B(x,\l)\,d\l$,
  $\hk(x)=\tint B(x,\l)^\dag\s\tau(\d/\d\l,\l)\s B(x,\l)\,d\l$, with $\tau(\d/\d\l,\l)\equiv -i\s\half\s\hbar\s\s[(\d/\d\l)\l+\l(\d/\d\l)]$. Here  $B(x,\l)\equiv A(x,\l)+c(\l)\mathbb{I}$, where $[A(x,\l),\s A(x',\l')^\dag]=\delta^s(x-x')\s\delta(\l-\l')\one$, $A(x,\l)$ annihilates the `no-particle' state $|0\>$, i.e., $A(x,\l)\s|0\>=0$ for all arguments, and $c(\l)$ is the real `model function' (defined below). Local products  are formally given, for example, by
   \bn &&\hph(x)\s\hph(x')=\tint\s B(x,\l)^\dag\,\l\,B(x,\l)\,d\l\s\cdot \tint\s B(x',\l')^\dag\,\l'\,B(x',\l')\,d\l'  \no\\
        &&\hskip4.56em=\tint\s\tint\s B(x,\l)^\dag\,\l\,[\s B(x,\l),\s B(x',\l')^\dag\s]\,\l'\,B(x',\l')\,d\l\,d\l' \no  \\
        &&\hskip15em +!\,\hph(x)\s\hph(x')\,!\\
        &&\hskip4.56em=\delta^s(x-x')\,\tint \s B(x,\l)^\dag \,\l^2\,B(x,\l)\,d\l+!\,\hph(x)\s\hph(x')\,! \;, \no \en
        where $!(\cdot)!$  denotes `normal ordering' for $B^\dag$ and  $B$, and which is now re-scaled (sometimes denoted by $R$ for `renormalized') by
        first letting $\widehat{\delta}$ denote a `smoothed out $\delta$-function' (e.g., a tall and narrow Gaussian function) and introducing
        the command `R$\delta$' meaning `restore $\delta$-functions'. It follows that  
          \bn &&\hskip-2em\hph_R^2(x)\equiv {R\delta}\s\lim_{x'\ra x} b\,\widehat{\delta}^s(0)^{-1}\,[\,  \widehat{\delta}^s(x-x')\,\tint B(x,\l)^\dag \,\l^2\,B(x,\l)\,d\l+!\,\hph(x)\s\hph(x')\,!\, ]\no\\
               &&\hskip0.9em\equiv b\tint B(x,\l)^\dag\s \l^2\s B(x,\l)\,d\l \;  \equiv\, \hph^2(x)\;\equiv \underline{\hph(x)\hph(x)} \;\not\equiv \hph(x)^2 \;, \en
                where $b$ is a positive factor with dimensions $(\rm length)^{-s}$, and, for simplicity, we will sometimes implicitly choose $b=1$.
                Note well the meaning of the several different expressions!

                The quantum Hamiltonian $\mfH$ is given by
     \bn  &&\mfH= \tint d^s\!x\s\tint\{B(x,\l)^\dag\,\mfh(\d/\d\l,\l)\, B(x,\l)\} \,d\l \no \\
    &&\hskip1.15em =\tint d^s\!x\s\tint\{A(x,\l)^\dag\s \mfh(\d/\d\l,\l)\s A(x,\l)\} \,d\l \;, \en
    a relation that requires
    \bn \mfh(\d/\d\l)\s\s c(\l)=0 \;;  \label{too}\en
moreover, to ensure that the ground state $|0\>$  is unique we require that $\tint c(\l)^2\,d\l=\infty$.
Guided by the classical Hamiltonian (3), we choose
   \bn &&\mfh(\d/\d\l,\l)= -\half\s\hbar^2 [\l(\d/\d\l)+\half]\s\l^{-2}\s[\l(\d/\d\l)+\half] + \half\s m_0^2\s \l^2+g_0\s \l^4 \no \\
       &&\hskip5em   = \half[- \hbar^2 (\d^2/\d\l^2) + \threequarter \hbar^2\s \l^{-2} + m_0^2\s \l^2]+g_0\s \l^4\;.  \label{two}    \en
The two lines of equation (\ref{two}) are elementary examples of two different approaches to express the Hamiltonian: the top line
`hides' the quantum correction, while the bottom line `shows' the quantum correction. In later sections we will encounter similar cases where
the Hamiltonian of more complex systems may also be presented in different fashions.

The model function $c(\l)$ that solves (\ref{too}) has a `large $\l$ behavior' and a `small $\l$ behavior' which leads to a functional behavior given by
$c(\l)=e^{-U(\l)/2}\s|\l|^{-1/2}$ for some function $U(\l)$ for which $-\infty<U(\l)<\infty$.
The `small $\l$ behavior' ignores the classical potential terms, and we can find the `small $\l$ behavior'
simply by observing that $[\l(\d/\d\l)+\half]\s F(\l)=0$ implies $F(\l)^{-1}\s\d\s F(\l)/\d\l=-1/(2\l)$, which leads to $F(\l)\propto \l^{-1/2}$.

The form of $c(\l)$ hints at the form of the Schr\"odinger ground-state wave function, namely
$ \Psi(\phi) = e^{-V(\phi)/2} \Pi_x|\phi(x)|^{-1/2} $, $-\infty< V(\phi)<\infty$, while one form of Schr\"odinger's equation is given by
\bn &&\hskip-2.8em i\s\hbar\s \d\s \Psi(\phi,t)/\d t = \{\tint[\half\s{\hk(x)\s \phi(x)^{-2}\hk(x)} + \half m_0^2 \phi(x)^2  
 + g_0 \phi(x)^4\s]\s d^s\!x\}\Psi(\phi,t), \en
  where
  \bn \hk(x)=-i\s\half\s\hbar\s[\p(x)(\delta/\delta\p(x))+(\delta/\delta\p(x))\p(x)]\;, \label{john}  \en
  which leads to a solution of the indicated form above.
However, these expressions are formal and $\Psi(\phi)$ is not square integrable as it stands. To rectify that we reintroduce the finite spatial lattice with $N'<\infty$ points used above to regularize ($r$) the ground state as $\Psi_r(\phi)= \Pi_\bk\s e^{-W(\phi_\bk)/2}\s ( ba^s)^{1/2}\s |\phi_\bk|^{-(1-2ba^s)/2} $, where we have added the dimensionless factor $ba^s$ needed to render $\Psi_r(\phi)^2$ effectively normalized in the regularized form.

 At this point it is worth mentioning that this form of the ground state allows numerous expectation values to be {\it finite}. To see this property, consider the expectation of numerous moments of the ground-state distribution given, with $p$ a positive integer, by
        \bn \<\s(\Sigma_\bl\hph_\bl^2)^{p}\>=\tint (\Sigma_\bl\phi_\bl^2)^p\, \Pi_\bk\s e^{-W(\phi_\bk)}\s ( ba^s)\s |\phi_\bk|^{-(1-2ba^s)}\;d\phi_\bk\;. \en
       Now introduce hyper-spherical coordinates $\phi_\bk\equiv \rho\s\eta_\bk$, $\rho\ge0$, $-1\le\eta_\bk\le1$, where $\rho^2=\Sigma_\bk \phi_\bk^2$ and $1=\Sigma_\bk
\eta_\bk^2$, which leads to
    \bn &&\<\s(\Sigma_\bl\hph_\bl^2)^{p}\>=\tint \rho^{2p}\, [\Pi_\bk\, e^{-W(\rho\eta_\bk)}\s ( ba^s)\s |\eta_\bk|^{-(1-2ba^s)}\s] \no \\   &&\hskip6em \times 2\s\delta(1-\Sigma_\bk\eta_\bk^2)\,[\Pi_\bk \s d\eta_\bk\s]\;\rho^{(R-1)}\,d\rho\;,\en
      where the usual measure factor $\rho^{(N'-1)}$ is effectively changed to $\rho^{(R-1)}$, with $R\equiv 2ba^s\s N'<\infty$ for a bounded spatial volume.
      This effective change of the power, i.e., $\rho^{(N'-1)}\rightarrow \rho^{(R-1)}$, in multi-field integrals eliminates general divergences that would normally arise as $N'\rightarrow\infty$, and that elimination, which will arise again in later sections, occurs for a special form of the `small field behavior' of suitable eigenfunctions.\footnote{The role of a `small field behavior' is also addressed in Chaps.~9 \& 10 of \cite{jrkEQ}.}

The characteristic function for these models takes the form
\bn  &&C(f)=\lim_{a\ra0}\s\Pi_\bk\s\tint \,\{ e^{ i f_\bk \phi_\bk/\hbar}\,e^{-W(\phi_\bk)}\, ( ba^s)\,|\phi_\bk|^{-(1-2ba^s)}\s\}\, d\phi_\bk \no\\
     &&\hskip2.48em   =\lim_{a\ra0}\s\Pi_\bk\,\{1 - (ba^s)\tint [1- e^{ i f_\bk \phi_\bk/\hbar} ]\,\,e^{-W(\phi_\bk)} \,|\phi_\bk|^{-(1-2ba^s)\,}d\phi_\bk \}\no\\  &&\hskip2.48em =\exp\{-b\tint d^s\!x\s\tint[1-e^{if(x)\s\l/\hbar}] e^{-w(\l,\hbar)}\,d\l/|\l|\}\;, \label{tyu} \en
     where $\phi_\bk\ra\l$, and $w$ may involve parameter renormalization as well. The result is the only other outcome of the Central Limit Theorem, namely, a
     (generalized) Poisson distribution. It is noteworthy to observe that the factor $\threequarter$ in (\ref{two}) is a special fraction
      that leads to the
      well-defined limit in (\ref{tyu}) as $a\ra0$ and that fraction was a direct result of adopting an affine quantization and not a canonical
      quantization. Moreover,
      the limit of such models as $g_0\ra0$ is {\it not} the free model (Gaussian) but becomes a pseudofree model (in this case, Poisson) \cite{jrk1}.

     It is straightforward to study the imaginary-time propagator for these models. As customary, the initial functional integral  has the form
       \bn && K(\phi'',T;\phi',0) =\<\phi''|\s e^{-{\mfH}\s T/\hbar}\s|\phi'\>     \\
       &&\hskip1em
      = {\cal M}\int e^{-{\t\frac{1}{\hbar}}\tint_0^T\,\tint \{ \half [\dot{\phi}(x,t)^2+m_0^2\s\phi(x,t)^2  ]+g_0\s \phi(x,t)^4\}\,
       d^s\!x\,dt}\;{\cal D}\phi \;, \no \en
      which is formal. To give it proper meaning, we choose the limit of a regularized functional integral, with an $\hbar$-counterterm,  as given by
        \bn && K(\phi'',T;\phi',0)=\<\phi''|\s e^{-{\mfH}T/\hbar}\s|\phi'\>    \\
        &&\hskip1em =\lim_{a\ra0,\delta\ra0} M_{a,\delta} \int \, e^{-{\t\frac{1}{\hbar}}{\t\Sigma_{\bk,k_0}}
    \{  \half [  (\phi_{\bk,k_0+1}-\phi_{\bk,k_0})^2\delta^{-2} +m_0^2\s\phi^2_{\bk,k_0} }     \no \\
     &&\hskip5em
        ^{{\t+\hbar}^2\s
        (\half-ba^s)(\threebytwo-ba^s)\s\phi^{-2}_{\bk,k_0}\s a^{-2s}] +g_0\s \phi^4_{\bk,k_0}   \}
           \,a^s\s\delta }   \;{\t \Pi}_{\bk,k_0}\s d\phi_{\bk,k_0}\;. \no \en

     Several features of these models are interconnected and choosing one of them may lead to knowledge of another one. For example, the
     ground state of a quantum system determines the Hamiltonian operator (up to a constant) which implicitly determines all there is to know
     about a given system. Based on that remark, we
     now focus on the regularized form of the ground state, or another suitable state, for the models to come and use that information
     to suggest regularized states for more complicated models.

\subsection{Classical/Quantum Connection}
Enhanced quantization favors different rules than those of canonical quantization  when seeking to pass from a quantum level to a classical level.
In the present case, we initially introduce an appropriate set of affine coherent states \cite{jrkEQ}, and, with $c\p(x)\ne0$ and dimensionless,
we choose the states
   \bn  |\pi,\p\>\equiv \exp[(i/\hbar)\tint \pi(x)\s\hph(x)\,d^s\!x]\,\exp[-(i/\hbar)\tint\s\ln(c|\p(x)|)\s\hk(x)\,d^s\!x]\,|\beta\>\;, \en
   which leads to the enhanced (since $\hbar>0$) classical Hamiltonian  given  by
     \bn &&H(\pi,\p)=\<\pi,\p|\,\mfH'(\hk,\hph)\,|\pi,\p\>
            =\<\beta|\,\mfH'(\hk+ \pi\s c|\p|\,\hph, c|\p|\,\hph)\,|\beta\>\;. \en
As an example, if $\mfH'(\hk,\hph)=\tint\,[\s\half\s\underline{ \hk(x)\,\hph^{-2}(x)\,\hk(x)}+\half\,m^2_0\,\hph^2(x)+g_0\,\hph^4(x)\s]\;d^s\!x$, along   
with $\<(\cdot)\>\equiv\<\beta|(\cdot)|\beta\>$ and a proper $|\beta\>$ and $c$ so that $c^2\<\s\hph^2(x)\s\>=1$ and $c^4\<\s\hph^4(x)\s\>=1+{\cal O}(\hbar)$, then
   \bn &&H(\pi,\p)=\tint \{ \half\,[\pi(x)^2 + m^2_0\,\p(x)^2]\,c^2\<\hph^2(x)\> \no \\ &&\hskip8em +g_0\,\p(x)^4\s c^4\<\hph^4(x)\>
   +{\cal O}(\hbar; \pi,\p)\s\} \,d^s\!x\;. \en

\section{Covariant Scalar Models in \\High Spacetime Dimensions}
\subsection{Standard Approach to the \\Quantum Formulation}
     The classical models (with ``c" denoting covariant) discussed in this section have a classical Hamiltonian given by
       \bn  H_c(\pi,\phi)=\tint\{ \half[\pi(x)^2+(\overrightarrow{\nabla}\phi)(x)^2+m_0^2\s\phi(x)^2]+g_0\s\phi(x)^4\}\;d^s\!x\;, \en
       which is just the classical ultralocal Hamiltonian plus a spatial gradient term. Such models are perturbably nonrenormalizable in 5 or more spacetime dimensions
 \cite{RUSIA,BCQ}, 
      and we primarily  focus on such models.  Canonical quantization of these models leads to a free (Gaussian) result \cite{AA,FF}, and
       we propose to use affine field variables for these models. The covariant Hamiltonian in affine variables is much like the ultralocal Hamiltonian in affine variables. 
       Specifically, we have
       \bn  H'_c(\k,\phi)=\tint\{ \half[\k(x)\phi(x)^{-2}\k(x) +(\overrightarrow{\nabla}\phi)(x)^2
       +m_0^2\s\phi(x)^2]+g_0\s\phi(x)^4\}\, d^s\!x\s. \en


         The principal difference between the covariant models and the ultralocal models is that the former involve spatial continuity while the latter does not. However, rather like the ultralocal ground state, the covariant ground state has a
       `large field behavior' and a `small field behavior' that respect the continuity. Although a formal affine quantization again leads to a
       similar counterterm as the ultralocal model, for the covariant scalar model we choose a different regularization, one that is not acceptable in the
       ultralocal case. Specifically,  we propose that the spatially regularized covariant
       scalar ground state is given by \cite{jrkJ}
            \bn \Psi_{c\s r}(\phi)=e^{-Y(\phi)/2}\s\Pi_\bk[\Sigma_\bl\s J_{\bk,\bl}\s\phi_\bl^2]^{-(1-2ba^s)/4} \;, \en
            where $J_{\bk,\bl}=1/(2s+1)$ for $\bl=\bk$ and $\bl$ is one of the $2s$ nearest-neighbor spatial [{\it sic}] points closest to the field
            at point $\bk$; otherwise
            $J_{\bk,\bl}=0$. The choice of the sum of a limited number of nearest-neighbor factors provides an escape
            from the ground-state distribution becoming a
            Poisson distribution even in the continuum limit; it also offers a perturbation procedure about the chosen ground state in which every
            term is finite (see, e.g., \cite{MM,jrk21}).


  Based on using the special counterterm for such models, the lattice regularized Hamiltonian operator for interacting models is chosen,
  with primed summation symbols meaning strictly a spatial sum, to be\footnote{The procedures for a quartic interaction also hold for other
  nonrenormalizable models with higher powers for their interaction.}
      \bn \H_{c\s r}\hskip-1.1em&&=-\half\hbar^2\s a^{-2s}{\t\sum'}_\bk\,\frac{\d^2}{\d\p_\bk^2}\,a^s+\half{\t\sum'}_{\bk,\bk^*}(\p_{\bk^*}-\p_\bk)^2\s a^{s-2} \no\\
    &&\hskip2em+\half\s m_0^2{\t\sum'}_\bk\p_\bk^2\,a^s+
    g_0{\t\sum'}_\bk\p_\bk^4\,a^s
    +\half\hbar^2{\t\sum'}_\bk\s {\cal F}_\bk(\p)\,a^s-E_0\;, \label{eH}\en
    where $\bk^*$ is one positive step forward from the site $\bk$ for each of the $s$ nearest lattice sites, in which the site labels may be  spatially  periodic.       
    In this expression, the counterterm is proportional to $\hbar^2$, and specifically is chosen so that
    \bn {\cal F}_\bk(\p)\hskip-1.1em&& \equiv \frac{a^{-2s}}{{\Pi_\bl}[\Sigma'_{\bm}\s J_{\bl,\bm}\s \phi_\bm^2]^{-(1-2ba^s)/4}}\,
    \frac{\d^2\,\Pi_\bl\s[\Sigma'_{\bm}\s J_{\bl,\bm}\s \phi_\bm^2]^{-(1-2ba^s)/4}} {\d\s\phi^2_\bk}\no \\
    &&\hskip0em=\quarter\s(1-2ba^s)^2\s
          a^{-2s}\s\bigg({{\ts\sum}'_{\s \bl}}\s\frac{\t
  J_{\bl,\s \bk}\s \p_\bk}{\t[\Sigma'_\bm\s
  J_{\bl,\s \bm}\s\p_\bm^2]}\bigg)^2\no\\
  &&\hskip.5em-\half\s(1-2ba^s)
  \s a^{-2s}\s{{\ts\sum}'_{\s \bl}}\s\frac{\t J_{\bl,\s \bk}}{\t[\Sigma'_\bm\s
  J_{\bl,\s \bm}\s\p^2_\bm]} \no\\
  &&\hskip1em+(1-2ba^s)
  \s a^{-2s}\s{{\ts\sum}'_{\s \bl}}\s\frac{\t J_{\bl,\s \bk}^2\s\p_\bk^2}{\t[\Sigma'_\bm\s
  J_{\bl,\s \bm}\s\p^2_\bm]^2}\;. \label{eF} \en
  Although ${\cal F}_\bk(\p)$ does not depend only on $\p_\bk$, it nevertheless becomes a local potential
  in the formal continuum limit.

\subsection{Non-standard Approach to the \\Quantum Formulation}
 The foregoing analysis involves one form of regulation which converges to the correct formulation as the regulation is removed, i.e., as $a\rightarrow 0$.
 However, there are other approaches that have different regularization procedures but still lead to the correct formulation as $a\rightarrow 0$. One
 alternative procedure is highly worth discussing since it offers a simple and more natural overall procedure. In particular,
  rather than (\ref{eH}), we can choose the regularized, Schr\"odinger representation, Hamiltonian operator given by
  \bn \H_{c\s r}\hskip-1.1em&&=\half\s{\t\sum'}_\bk\,\hk_\bk\,[\Sigma'_\bl J_{\bk,\bl}\s\s\p_\bl^2]^{-(1-2ba^s)}\,\hk_\bk\,a^s+\half{\t\sum'}_{\bk,\bk^*}(\p_{\bk^*}-\p_\bk)^2\s a^{s-2}   \no\\
   &&\hskip2em +\half\s m_0^2{\t\sum'}_\bk\p_\bk^2\,a^s+
    g_0{\t\sum'}_\bk\p_\bk^4\,a^s-E'_0\;, \label{eH2}\en
    where $\hk_\bk=-i\s\half\s\hbar\s[\s\p_\bk\s(\d/\d\p_\bk)+ (\d/\d\p_\bk)\s\p_\bk\s]\s\s a^{-s}$. This alternative expression incorporates suitable $\hbar$-additions as the regularization is removed. Observe that the expression in (\ref{eH2}) is {\it a natural transition from the classical Hamiltonian to the quantum Hamiltonian.} Such a property will serve us well when we seek the quantization of the Hamiltonian for general relativity.

\subsection{Smaller Spacetime Dimensions}
  We emphasize that we may build a regularized functional integral to determine the imaginary-time propagator for the covariant scalar models much like
   (12) for the ultralocal model. This possibility leads us to important studies.
   It is known that Monte Carlo (MC) calculations of the canonical quantization of the perturbably {\it renormalizable}, quartic, covariant scalar field in 4 spacetime dimensions effectively point to
   a free quantum theory \cite{fw}. The proposed quantization procedures of this section, including (\ref{eH}), have also been extended to 4 spacetime dimensions, and a preliminary
   MC study with the novel counterterm  points toward a positive renormalized coupling constant, potentially becoming a non-trivial result \cite{js}. Although this initial study had to stop too early, there is presently a new study that hopefully will resolve this issue.

   Moreover, one can extend the models of this section to spacetime dimension $n=3$ and $n=2$. Conventional quantization  also provides
   acceptable quantum solutions
   for such models as $\phi^4_2$ and $\phi^4_3$ \cite{JJ}, but not for higher $n\ge4$. However, the extended models of this section can offer
   alternative solutions to those generated by canonical procedures, and one can also consider `mixed models' of the kind $g\s\phi^4_3+g'\s\phi^8_3$ as well defined theories when $g\ge0$ and $g'\ge0$ are varied  arbitrarily \cite{jrk21}.

\section{Quantum General Relativity}

\subsection{ Without Constraints: Why No Constraints}

In this subsection we consider general relativity with fixed terms that are usually treated as constraints, and which will be properly dealt with below.
The reader may well ask why we ignore the constraints. The reason is that there are choices to be made: Dirac favors quantizing first
and applying the constraints second, while others choose to enforce constraints before quantizing.

Consider the toy example where
the classical action functional is given by
   \bn A_1=\tint\{p(t)\s\dot{q}(t)-\l(t) [p(t)^2+q(t)^4-E]\s\}\,dt \en
 and the question proposed is: ``What values of $E$ lead to valid quantum stories?''
 Here $\l(t)$ denotes a Lagrange multiplier to enforce the classical constraint $p(t)^2+q(t)^4=E$. Quantizing first offers the values $\{E_n\}$, $n=1,2,\cdots$, for which there is a non-zero vector $
|n\>$, where $(P^2+Q^4)\s|n\>=E_n\s|n\>$. Hence, we have a correct answer when we quantize first, but there is no such solution if the constraint is satisfied before quantization. Moreover, the simple example given by
  \bn A_2=\tint [ p(t)\s\dot{q}(t)-\l_1(t)\s p(t)-\l_2(t)\s q(t)\s]\,dt\;, \label{21}\en
     with the Poisson bracket $\{q(t),\s p(t)\}=1$, has two Lagrange multipliers that lead to the second-class constraints, $p(t)=0$ and $q(t)=0$. Moreover, the zero classical constraints can not lead to zero
  quantum constraints, i.e., $P=0$ and $Q=0$, since $[Q,\s P]=i\hbar\one$. The best that can be achieved, for example, is a projection
  operator\footnote{Note: The $\delta$-function here is {\it not} the Dirac $\delta$-function!}
    $\mathbb{E}(P^2+Q^2\le \delta(\hbar)^2)$   where $\hbar\le\delta(\hbar)^2<3\hbar$ which encompasses just one state, $\mathbb{E}=|0\>\<0|$, and points toward  a single Hilbert space vector, a result offered by the projection operator method of dealing with all constraints \cite{pom},
   which we will exploit below.

   These examples explain why we choose to quantize first and then enforce the quantum constraints carefully to account for possible second-class constraints
   (which quantum gravity is known to possess).

\subsection{The Gravitational Hamiltonian}
Using the ADM  phase-space variables \cite{adm}, the classical Hamiltonian is given, for $a,b=1,2,3$, and assumed summation of index pairs, by
   \bn H(\pi,g)=\!\int{\big\{}\s\frac{1}{\sqrt{g(x)}}\s[\pi^a_b(x)\s\pi^b_a(x)-\half\s\pi^a_a(x)\s\pi^b_b(x)]+\sqrt{g(x)}\,^{(3)}\!R(x)
   \,{\big\}}\;d^3\!x\s,   \label{affine}\en
where $\pi^a_b(x)\equiv \pi^{ac}(x)\s g_{bc}(x)$,  the Poisson bracket   $\{g_{ab}(x),\s\pi^{cd}(x')\}$ $=\delta^3(x-x')\,\half[\delta^c_a\delta^d_b+\delta^c_b\delta^d_a]$, and $^{(3)}\!R(x)$ is the scalar curvature
in the $3$ spatial coordinates. The physics of the metric requires that
$g(x)\equiv \det[g_{ab}(x)]>0$ for all $x$.
When quantized, this positive metric requirement implies  that the
momentum variables can not be made locally self adjoint, and this makes canonical quantization especially difficult. Happily, affine quantization
can come to the rescue \cite{jrk22}!

Instead of promoting $g_{ab}(x)$ and $\pi^{cd}(x)$ to quantum operators, we promote the metric tensor $g_{ab}(x)$ and the momentric tensor $\pi^c_d(x)\,[\equiv\pi^{ca}(x)\s g_{da}(x)]$, with the variable $\pi^c_d(x)$ awarded the special name as the `momentric field', a name
derived from its {\it momen}tum and me{\it tric} fields. {\it Observe, then, that the classical Hamiltonian in (\ref{affine})     
is already expressed in affine variables!}

Fortunately, it  turns out that the two variable sets, $g_{ab}(x)$ and $\pi^c_d(x)$,
have a closed algebra, which we already express in the form of commutators of the
local field operators, specifically
 \bn   &&[\hp^a_b(x),\s \hp^c_d(x')]=i\s\half\,\hbar\,\delta^3(x-x')\s[\delta^a_d\s \hp^c_b(x)-\delta^c_b\s \hp^a_d(x)\s]\;,    \no \\
       &&\hskip-.10em[\hg_{ab}(x), \s \hp^c_d(x')]= i\s\half\,\hbar\,\delta^3(x-x')\s [\delta^c_a \hg_{bd}(x)+\delta^c_b \hg_{ad}(x)\s] \;,      \\
       &&\hskip-.20em[\hg_{ab}(x),\s \hg_{cd}(x')] =0 \;. \no  \en
        In this case the operators are given by
      \bn  &&\hg_{ab}(x)=\tint_+ B(x, \gamma)^\dag\,\gamma_{ab}\,B(x,\gamma)\;d\gamma\;,   \\
            &&\hskip.3em\hp^c_d(x)=-i\s\half\s\hbar\,\tint_+\s B(x,\gamma)^\dag\s[\s\gamma_{dj}\s(\d/\d\gamma_{cj})+(\d/\d\gamma_{cj})\,\gamma_{dj}\s]\s B(x,\gamma)\,d\gamma\;,  \no \en
            where  $d\gamma=\Pi_{a\le b} \,d\gamma_{ab}$ and $\tint_+$ limits the range of integration to  $\{\gamma_{ab}\}>0$.
   Just as the case in Sec.~2.2, there are two irreducible representations of the metric tensor operator: one where the matrix $\{\hg_{ab}(x)\}>0$, which we accept, and one where the matrix $\{\hg_{ab}(x)\}<0$, which we reject.  When smeared with suitable
   test functions, the result is that both the metric and the momentric tensors can be self-adjoint operators, and the metric operators will satisfy the required positivity requirements!


     Note that
    we now reserve $g$ for $\det[g_{ab}]$ and introduce $\{g\}$ for the $3\times3$ general elements of the metric tensor.
    The Schr\"odinger representation of the proper Hilbert space vectors  is then given by $\Psi(\{g\})$.
   We accept the fact that the Schr\"odinger representation of eigenfunctions of the Hamiltonian operator have a `large field behavior' and
   a `small field behavior', as suggested by the previous discussion, and the Hamiltonian operator eigenfunctions are formally given by $\Psi(\{g\})=W(\{g\})\,[\Pi_x g(x)^{-1/2}]$, where the `small field behavior' is formally obtained by the relation $ \hp^a_b \s F(g)=0$, which implies that
   $[g_{bc}\s (\d/\d g_{ac})+\half\delta^a_b]\s F(g)=0$ and this leads to $g_{bc}\s g^{ac}\s g\s\s dF(g)/d g + \half\delta^a_b\s F(g)=0$, which requires that
   $ g\s d\s F(g)/d g+\half\s F(g)=0$; hence $F(g)\propto g^{-1/2}$.

   The reader will note that the factor $g(x)^{-1/2}$ differs from the traditional factor $g(x)^{1/2}$ which is used in expressions like
   $g(x)^{1/2}\,d^3\!x$
   to serve as a scalar. A scalar is also formally given by $ \delta^3(x)\,d^3\!x$ as well. In fact, the term $g(x)^{-1/2}$ normally appears with two
   factors of $\delta^3(0)$--e.g., two terms from (\ref{john})--which becomes $[\delta^3(0)\,\delta^3(0)]g(x)^{-1/2}\,d^3\!x$, and this is also a formal scalar. In regularized form, $\delta^3(0)$ is represented by $a^{-3}$,
   $d^3\!x$ by $a^3$, and $g(x)$ by $g(\bk a)=g_\bk$.\footnote{While we have focused on quantum gravity in 4 spacetime dimensions, it is clear that
   our analysis can also be used in other spacetime dimensions as well.}

\subsubsection{Hamiltonian-free quantization}
   At this juncture the analysis points toward a theory known as `Strong Coupling Quantum Gravity'
   \cite{PP,PP2} in which the scalar curvature $^{(3)}\!R(x)$
   in Eq.~(\ref{affine}) is dropped and replaced by the cosmological constant $\Lambda$ resulting in a new Hamiltonian having no spatial derivatives, which
   thus becomes an ultralocal model; this particular model
   is justified as a possible starting point for a perturbation analysis that reintroduces the scalar curvature. For such a problem the `small field
   behavior' is formally given by $\Pi_x\s g(x)^{-1/2}$ and regularized by the expression $\Pi_\bk (ba^3)^{1/2}\s g_\bk^{-(1-ba^3)/2}$, rather like the
   earlier expressions for ultralocal models. The measure $\Pi_x\s \Pi_{a\le b}\s dg_{ab}(x)$ is regularized and becomes\footnote{This measure is discussed on page 205 in \cite{jrkEQ}}  $\Pi_{\bk,\s a\le b}\s dg_{\bk,\s ab}$.
    However, we choose not to pursue `ultralocal general relativity', and we return to the model in (\ref{affine}) which is not an ultralocal model.

\subsubsection{Restoring the Hamiltonian}
    As were the procedures in Sec.~3, we regularize the chosen eigenfunctions  by replacing the spacial continuum by a set of $N'<\infty$ points labeled by the usual points $\bk a$ and introduce a regularized ($r$) eigenfunction given by
   \bn  \Psi_r(\{g\})=W_r(\{g\})\,\{\Pi_\bk\,(ba^3)^{1/2}\,[\Sigma_\bl J_{\bk,\bl}\,g_\bl]^{-(1-ba^3)/2} \}\;, \en
  where the factors $J_{\bk,\bl}$ are the same factors as in Sec.~3.1.  Because the affine variable complex in (20) is not positive definite, the quantum eigenvalues will, most likely,
    range over the whole real line. Thus, $W_r(\{g\})$ will, again most likely, be positive and negative for all eigenfunctions, and we focus attention on an
    appropriate eigenfunction that is nonzero in the vicinity of very small values of $g$. Just as in the covariant scalar case, we choose the `large field behavior' of the regularized quantum Hamiltonian operator from the classical Hamiltonian, and we choose the `small field behavior' of the
  regularized quantum Hamiltonian,
   specifically, as the factor $\Pi_\bk[\Sigma_\bl J_{\bk,\bl}\,g_\bl]^{-(1-ba^3)/2}$. Based on Sec.~3.2, we are led to the regularized form of the quantum Hamiltonian in the Schr\"odinger representation given by
      \bn \mfH_r= {\t\sum}'_\bk \{ \, {\hp^a_{b\s\bk}\s\s {\bf J}_\bk(g_\bk)\s \hp^b_{a\s\bk}} -\half\s{\hp^a_{a\s\bk}\s\s {\bf J}_\bk(g_\bk) \s\hp^b_{b\s\bk}}
      +g_\bk^{1/2}\,^{(3)}\!R_\bk +g^{1/2}_\bk\s \Lambda \} \,a^3\;,  \label{eK}\en
      where ${\bf J}_\bk(g_\bk)\equiv[\Sigma_\bl J_{\bk,\bl}\,g_\bl]^{-(1-ba^3)/2} $, and
       \bn  &&\hp^a_{b\s \bk}=-i\s\,\half\s\hbar\{\frac{\t\d}{\t\d\s g_{ac\s\bk}}\s\,g_{bc\s\bk}+ g_{bc\s\bk}\,\frac{\t\d}{\t\d\s g_{ac\s\bk}}\} \s a^{-3}\;.\label{eJ}\en


\subsection{Enforcing the Constraints}
The classical action functional for gravity is given \cite{adm} by
  \bn A=\tint\tint\,\{ \pi^{ab}(x,t)\s {\dot g}_{ab}(x,t) - N^a(x,t)\s \pi^b_{a\s |b}(x,t)-N(x,t)\s H(x,t)\,\}\, d^3\!x\,dt\;, \en
  where the Lagrange multipliers, the lapse, $N(x,t)$, and the three shifts, $N^a(x,t)$, enforce the classical Hamiltonian  constraints, $H(x,t)=0$,
  and the classical diffeomorphism constraints, $\pi^b_{a\s |b}(x,t)=0$, for all $x\,\&\, t$. Since the classical constraints are first class, the Lagrange multipliers can assume any
  values in the equations of motion, such as $N(x,t)=1$, etc. However, in the quantum theory, $H(x,t)$ and $\pi^b_{a\s |b}(x,t)$ become operators, while
  $N(x,t)$ and $N^a(x,t)$ remain classical functions.

  Let us focus on the regularized classical Hamiltonian constraints, $H_\bk=0$, for all $\bk$, and the three regularized classical diffeomorphism
   constraints, $\pi^a_{b\s \bk}$ $\!\! _{|\,a}$ $=0$, for all $b$ and $\bk$,  where $_|$ denotes a regularized
   covariant scalar derivative. The four regularized quantum constraints should follow the classical story as closely as possible, and so, following Dirac,
    we initially propose that vectors in the physical Hilbert space obey
    $\mfH_\bk\s|\Psi\>_{phys}=0$ for all $\bk$ and $\hp^a_{b\s \bk}$ $\!\! _{|\,a}\s|\Psi\>_{phys}=0$ for all $b$ and $\bk$, for a `wide class' of non-zero Hilbert space vectors.
    However, that goal is not possible since, for certain $\bk$ and $\bm$, $[\mfH_\bk,\s\mfH_\bm]\s|\Psi\>_{phys}\ne0$ due to
    quantum second-class constraints. Instead, we choose an appropriate projection operator $\mathbb{E}=\mathbb{E}(N'^{-1}\s[\s\Sigma_\bk\,\mfH^2_\bk+\Sigma_{a,\s\bk}\,\hk^{b\,2}_{a\s\bk|b}\s]\,\le\,\delta(\hbar)^2\,)$, which is adjusted so
    that the constraints have the smallest, non-vanishing values.  If $\<\Psi|\Phi\>$ denotes the inner product in the original, kinematical Hilbert space $\mathcal{H}$, then $\<\Psi|\,\mathbb{E}\,|\Phi\>$ denotes the inner product in the reduced, physical Hilbert space $\mathcal{H}_{phys}$; or symbolically stated, $\mathcal{H}_{phys}=\mathbb{E}\,\mathcal{H}$.

    The projection operator $\mathbb{E}$ can be constructed by a suitable functional integral \cite{proj,pom}.
    In the general case, choosing a set of arbitrary, self-adjoint, constraint operators, $C_\alpha$, where $\alpha\in\{1,2, ..., A\}$, we construct a
    functional integral given by
      \bn\mathbb{E}(\Sigma_\alpha C^2_\alpha\le \delta(\hbar)^2\,)=\int\, \mathbb{T}\, e^{-i\tint_0^T\,\Sigma_\alpha C_\alpha\,\l_\alpha(t)\,dt}\,\mathcal{D} R(\l)\;,\en
      where $\mathbb{T}$ implies a time-ordered integral and $R(\l)$  is a suitable weak measure (see \cite{proj}) which is dependent only on: (i) the time $T$, (ii) the upper limit $\delta(\hbar)^2\ge0$, and (iii) the number of constraints $A\le\infty$. {\it The measure $R(\l)$ is completely independent of the choice of the constraint
      operators $\{C_\alpha\}_{\alpha=1}^N$! }


\section{Summary}
\subsection{Other Quantum Gravity Studies}
    The classical theory of gravity, proposed by Einstein, is a remarkable and generally accepted theory. On the other hand, its quantum version has
    been proven to be not only difficult but actually impossible using the tools of canonical quantization, which have shown it to be perturbably
    nonrenormalizable, a traditional `death threat' to any theory. Various modifications of the fundamental dynamical equations have led to systems that
    lead to certain results, but invariably these results do not represent a valid quantization of general relativity. Our efforts aim to
    provide a valid quantization of general relativity, and for that purpose we use methods that encompass canonical quantization.

    Notably, the traditional rules of quantization, namely canonical quantization, have a weak spot that has recently been overcome in the form of enhanced quantization
    \cite{jrkEQ}, an improvement which can resolve various issues when canonical quantization fails.  The classical version of the ultralocal scalar models
    examined in Sec.~2, is well behaved classically, but becomes perturbably nonrenormalizable when studied by canonical quantization; yet it is an important example
    because the unusual $\hbar$-counterterm that leads to an acceptable quantum theory has been found among the vast set of possibilities \cite{JrK,BCQ}.
    More recently, it has been discovered that the correct counterterm automatically pops up in the realm of affine quantization, a branch of enhanced quantization.
    This happy coincidence has been featured in this article by discussing a set of more complex models each of which is  perturbably nonrenormalizable when
    analyzed by canonical  quantization procedures. Instead, when these models are analyzed by affine quantization procedures a particular form of
    an $\hbar$-counterterm automatically appears. While this feature is clearly correct for the ultralocal models, it is not yet clear whether this desirable
    feature extends to covariant scalar models in spacetime dimensions 5 and greater, nor for the gravitational models in spacetime dimension 4. To check the
    worthiness of these cases, especially the case of gravity, will most likely require Monte Carlo studies of certain properties for these models.
\subsection{Some Important Equations}
    The proper Hilbert space and related operators that our analysis features for gravity are summarized here. As noted before, the principal affine field operators are
        \bn  &&\hg_{ab}(x)=\tint_+ B(x, \gamma)^\dag\,\gamma_{ab}\,B(x,\gamma)\;d\gamma\;,   \\
            &&\hskip.3em\hp^c_d(x)=-i\s\half\s\hbar\,\tint_+\s B(x,\gamma)^\dag\s[\s\gamma_{dj}\s(\d/\d\gamma_{cj})+(\d/\d\gamma_{cj})\,\gamma_{dj}\s]\s B(x,\gamma)\,d\gamma\;,  \no \en
            where  $d\gamma=\Pi_{a\le b} \,d\gamma_{ab}$ and $\tint_+$ limits the range of integration to  $\{\gamma_{ab}\}>0$.

            Our regularized version of the metric operators is given by
            \bn \hg_{ab\,\bk}=\tint_+\, B_\bk(\g)^\dag\,\g_{ab}\,B_\bk(\g)\;d\g\;, \en
            and the spatial first derivative of the metric operator is  $[\hg_{ab\,\bk+\bl}-\hg_{ab\,\bk}]\,a^{-1}$, where $\bl$ represents
            a single step away from $\bk$ in a given spatial direction. Multiple derivatives follow a similar pattern. However, powers of metric derivatives
            are somewhat more unusual, and specifically the regularized version of $\hg_{ab,c}^2(x)$ is given by
              \bn [\hg_{ab\,\bk+\bl}-\hg_{ab\,\bk}]^2_R\s a^{-2} =\tint_+\,\{\,B_{\bk+\bl}^\dag(\g)\,\g_{ab}^2\,B_{\bk+\bl}(\g)
                \,-B_{\bk+\bl}^\dag(\g)\,\g_{ab}^2\,B_{\bk}(\g)   \no\\   \,-B_{\bk}^\dag(\g)\,\g_{ab}^2\,B_{\bk+\bl}(\g)    \,+B_{\bk}(\g)^\dag\,\g_{ab}^2\,B_{\bk}(\g)\,\} \,d\g\;a^{-2}\;. \en

      The Hamiltonian operator and the diffeomorphism operators are given by
        \bn &&\tint\tint_+\,\{\,B(x,\g)^\dag\,[\xi(\g)^a_b\s \det(\g)^{-1/2}\s\xi(\g)^b_a \no\\
        &&\hskip8em -\half\,\xi(\g)^a_a\s \det(\g)^{-1/2}\s\xi(\g)^b_b]\,B(x,\g) \no \\
          &&\hskip2em  + B(x,\g)^\dag\,[\, N(x,t)\s\det(\g)^{1/2}\,^{(3)}\!R(\g)   \\ 
          &&\hskip8em + N^a(x,t)\,\xi(\g)^b_{a\,_|\s b}\,]\, B(x,\g)\, \}\,d\g\,d^3\!x \;,\no\en
          where $\xi(\g)^a_b=-i\s\half\s\hbar\s[\g_{bc}\s(\d/\d\g_{ac})+(\d/\d\g_{ac})\s\g_{bc}]$.
\subsection{Future Analysis}
        It is straightforward to generate regularized versions of variables and suitable equations, and one may create a propagator as the limit of a functional integral,
          much like the ultralocal or (implicitly) the covariant scalar field story, except now there would be an additional functional integral to enforce
          the constraints using the projection operator method. Indeed, such a functional integration already appears in \cite{jrkEQ}, Sec.~10.2.2.
          Regularized versions of that particular functional integral, or alternative functional integrals such as might appear in \cite{pom},
          are, very likely, to serve as suitable candidates for Monte Carlo studies. Of course, these continuum integrals will need to be
          regularized. In this effort, the expressions (\ref{eH}) and (\ref{eF}), which seem relevant for covariant scalar fields, may lend credence
          to (\ref{eK}) and (\ref{eJ}) in any similar study of general relativity. To establish the similarity of these two classes of
          perturbably nonrenormalizable studies, and to have faith in the preliminary Monte Carlo evidence and the relevance of that possibility
          to play a similar role in gravity is the central message of this article. Hopefully, this potential connection may be examined by others
          and, possibly, they may justify the essence of this analysis.


\begin{thebibliography}{99}
\bibitem{jrk1} J.R. Klauder, ``{Quantum Field Theory With No Zero-Point Energy}'',   JHEP 05(2018)191; arXiv:1803.05823. 

\bibitem{JrK} J.R. Klauder, ``Ultralocal Scalar Field Models", Commun. Math. Phys. {\bf 18}, 307-318 (1970).   

\bibitem{jrkEQ} J.R. Klauder, {\it Enhanced Quantization: Particles, Fields \& Gravity}, (World Scientfic, Singapore, 2015).   

\bibitem{RUSIA}  O.A. Ladyzenskaja, V. Solonnikov,  and N.N. Ural'ceva, {\it Linear and            
     Quasi-linear Equations of Parabolic Type}, (Am. Math. Soc., Providence, Vol. 23, 1968).


\bibitem{BCQ} J.R. Klauder, {\it Beyond Conventional Quantization}, (Cambridge University Press,  
Cambridge, 2000 and 2005).

\bibitem{AA}  M. Aizenman, ``Proof of the Triviality of                       
$\varphi^4_d$ Field Theory and Some Mean-Field Features of Ising
Models for $d>4$'', { Phys. Rev. Lett.} {\bf 47}, 1-4, E-886
(1981).

\bibitem{FF} J. Fr\"ohlich, ``On the Triviality of $\l\varphi^4_d$             
Theories and the Approach to the Critical Point in $d\ge 4$
Dimensions'', { Nuclear Physics B} {\bf 200}, 281-296 (1982).


\bibitem{jrkJ} J.R. Klauder, ``Scalar Field Quantization Without Divergences In All Spacetime Dimensions'',     
J. Phys. A: Math. Theor. 44, 273001 (30pp) (2011); arXiv:1101.1706.



\bibitem{MM}  J.R. Klauder, ``Nontrivial Quantization of $\phi^4_n,\, n\ge2$'', Theor. Math. Phys. 182 (2015) no.1, 83-89;  
Teor. Mat. Fiz. 182 (2014) no.1, 103-111; arXiv:1405.0332.

\bibitem{jrk21} J.R. Klauder, ``Mixed Models: Combining Incompatible Scalar Models in any Spacetime Dimension'', 
Int.~J.~Mod.~Phys. A32 (2017) no.~01, 1750001; arXiv:1605.06725.



\bibitem{fw} B. Freedman, P. Smolensky, D. Weingarten, ``Monte Carlo Evaluation of the Continuum Limit of    
 $(\phi^4)_4$ and $(\phi^4)_3$'', { Phys. Lett.} 113B, 481-486 (1982).

\bibitem{js}  J. Stankowicz, private communication.   

\bibitem{JJ} J. Glimm and A. Jaffe, {\it Quantum Physics}, (Springer-Verlag,  New York, 2nd Edition, 1987).    

\bibitem{pom} J.R. Klauder, {\it A Modern Approach to Functional Integration}, (Birkhauser, Boston, MA, 2010).   

\bibitem{adm}  R. Arnowitt, S. Deser, and C. Misner, 
in  {\it Gravitation: An Introduction to Current Research}, Ed. L. Witten,
(Wiley \& Sons, New York, 1962), p. 227; arXiv:gr-qc/0405109.

\bibitem{jrk22} J.R. Klauder, ``Recent Results Regarding Affine Quantum Gravity'', J. Math. Phys. 53, 082501 (2012); arXiv:1203.0691.    

\bibitem{PP}  M. Pilati, ``Strong Coupling Quantum Gravity. 1. Solution In A Particular Gauge'',
 Phys. Rev. D 26, 2645 (1982).

\bibitem{PP2}  M. Pilati,  ``Strong Coupling Quantum Gravity. 2. Solution Without Gauge Fixing'', Phys. Rev. D 28, 729 (1983).

\bibitem{proj} J.R. Klauder, ``Universal Procedure for Enforcing Quantum Constraints'', Nuclear Phys. B {\bf 547},  
397-412 (1999).                





.


\end{thebibliography}
\end{document}